\documentclass[sn-mathphys-num]{sn-jnl}

\usepackage{graphicx}%
\usepackage{multirow}%
\usepackage{amsmath,amssymb,amsfonts}%
\usepackage{amsthm}%
\usepackage{mathrsfs}%
\usepackage[title]{appendix}%
\usepackage{xcolor}%
\usepackage{textcomp}%
\usepackage{manyfoot}%
\usepackage{booktabs}%
\usepackage{algorithm}%
\usepackage{algorithmicx}%
\usepackage{algpseudocode}%
\usepackage{listings}%

\usepackage{siunitx}
\DeclareSIUnit\angstrom{\text{\AA}}
\usepackage{physics}

\usepackage[color=orange!60,textsize=scriptsize]{todonotes}
\theoremstyle{thmstyleone}%
\theoremstyle{thmstyletwo}%

\theoremstyle{thmstylethree}%

\usepackage{xcolor}

\begin{document}

\title[Article Title]{Neutron Scattering studies of the density dependence of One-dimensional Quantum Fluids}


\author[1]{\fnm{Sutirtha} \sur{Paul}}\email{spaul19@vols.utk.edu}
\author[2]{\fnm{Rahul} \sur{Soni} }\email{rahulsoni.phy@gmail.com}
\author[2]{\fnm{Garfield} \sur{Warren}}\email{gtwarren@iu.edu}
\author[1,3]{\fnm{Adrian} \sur{Del Maestro}}\email{adrian.delmaestro@utk.edu}
\author*[2]{\fnm{Paul E.} \sur{Sokol}}\email{pesokol@iu.edu}

\affil[2]{\orgdiv{Department of Physics}, \orgname{Indiana University}, \orgaddress{\city{Bloomington}, \postcode{47405}, \state{IN}, \country{USA}}}

\affil[1]{\orgdiv{Department of Physics}, \orgname{University of Tennessee}, \orgaddress{\city{Knoxville}, \postcode{37996}, \state{TN}, \country{USA}}}

\affil[3]{\orgdiv{Min Kao Department of Electrical Engineering and Computer Science}, \orgname{University of Tennessee}, \orgaddress{\city{Knoxville}, \postcode{37996}, \state{TN}, \country{USA}}}

\abstract{In previous work, we demonstrated that $^4$He confined within a templated porous material preplated with a noble gas to enhance dimensional reduction exhibits one-dimensional behavior. Neutron scattering revealed that the excitations in confined $^4$He are qualitatively distinct from those in two- and three-dimensional superfluid helium, aligning with quantum Monte Carlo calculations. These simulations suggest that three solid layers form, with a one-dimensional liquid located at the center of the pore. Here, we present neutron scattering results as the filling fraction of the one-dimensional core is varied. We observe that the one-dimensional liquid always forms with a fixed density, even as the pore filling increases. We posit that this is due to the growth of the one dimensional extent of the liquid as the filling is increased, but its density remains unchanged.
}

\keywords{Porous materials, 1D confinement, Quantum liquid}

\maketitle
\section{Introduction}\label{sec1}

One-dimensional systems have captivated researchers interest for their stark departure from the behaviors exhibited in two and three dimensions \cite{Giamarchi2004}. Unlike their higher-dimensional counterparts, which can be analyzed through the lens of quasi-particles, one-dimensional systems defy this paradigm entirely. Here, the fundamental excitations are collective and are governed at low energies and long wavelengths by the universally applicable Tomonaga-Luttinger liquid (TLL) theory \cite{Tomonaga:1950ak, Luttinger:1963gd, Mattis:1965iq, Haldane:1981eh}.  Despite its significance in elucidating the behaviors of both bosons and fermions in 1D, experimental tests of the predictions of TLL theory have been performed in low density fermionic systems \cite{jompol2009probing}, such as nanowires and nanotubes, insulating magnets \cite{Lake:2005ho} or ultracold bosonic gases confined to one dimension
\cite{haller2010pinning,yang2017quantum}.

In the low-energy regime, the physics of a one-dimensional system can be encapsulated by the Luttinger model Hamiltonian:
\begin{equation}
    H=\frac{\hbar v}{2\pi}\int_{0}^{L}dx\left[\frac{1}{K_L}\pqty{\frac{\partial\phi}{\partial x}}^{2}+K_L\pqty{\frac{\partial\theta}{\partial x}}^{2}\right]
    \label{Hamiltonian}
\end{equation}
In this expression, the conjugate Bose fields $\phi(x)$ and $\partial_x \theta(x) / \pi$ represent phase  and density fluctuations, respectively. The velocity of sound is denoted by $v$, while $K_L$, the Luttinger parameter, is determined by the details of interparticle interactions and the density of the system. Small values of $K_L$ are associated with superfluid-like phases, while larger values indicate solid-like behavior, however all correlation functions are described by algebraic decay and there is no true long-range order. The 1D liquid displays notable characteristics for small momentum, such as a linear dispersion relation, a gapless excitation spectrum, and spin-charge separation. Some of these signatures have been observed in systems such as trapped atoms and quantum wires \cite{Senaratne:2022fg}.  Ultracold gases permit interaction tuning from weak to strong coupling
\cite{haller2010pinning,meinert2015probing} at low density with contact interactions.  Confined helium provides a complementary setting in which to study one-dimensional quantum liquids with realistic interatomic interactions.

Helium isotopes have long been recognized as exemplary systems for probing theories of strongly interacting quantum matter and phase transitions in bosons, fermions, and mixed-statistics systems\cite{Ceperley1986,Whitlock1979,Sosnick1989}.  These fluids exhibit many interesting quantum phenomena, such as Bose condensation\cite{Sokol1995} and Fermi liquid behavior\cite{Wilks1987}, which have been extensively studied and are of current interest due to their observation in other systems, such as  Bose condensation in dilute gases\cite{Anderson1995,Bradley1995}, and excitonic systems\cite{Lin1993} and dirty Boson systems\cite{Landsberg1995}. These liquids have also occupied a central role as a model system for the study of critical phenomena\cite{Wilson1975,Ahlers1978},  and even the birth of the Universe\cite{Hendry1994}.

Liquid helium confined in 1D offers an attractive model system for the study of the behavior of a TLL where the density can be varied continuously from the weakly interacting to the strongly interacting regimes. The challenge to such studies of helium is identifying a platform for confining helium in 1D.  Previous experiments have investigated $^4$He in nanofabricated membrane pores \cite{duc2015critical}, mesoporous silica channels \cite{wada2001helium, taniguchi2010superfluidity}, and conducting pathways in solid helium \cite{vekhov2012mass}. These studies reported thermodynamic and transport signatures of reduced dimensionality, motivating direct spectroscopic measurements of 1D dynamics and their density dependence.  The main challenge is confining helium to length scales on the order of the helium correlation length, which is on the sub-nanometer scale.

Recently, Del Maestro et. al.~\cite{del2022experimental} have developed a nanoengineering technique were unambiguous evidence of Luttinger liquid behavior in $^4$He could be observed.  They utilized MCM-41, a templated porous glass with nanometer scale 1D pores.  While the as synthesized pores in MCM-41 are too large to effectively confine the helium to 1D,  preplating the pores with argon at high temperatures allows the pore diameter to be reduced significantly and TLL behavior can be observed using neutron scattering.

In this paper, we describe inelastic neutron scattering studies of one-dimensional $^4$He as a function of the core liquid density at the pore center. We employ the experimental platform established in Ref.~\cite{del2022experimental} utilizing MCM-41 nanopores preplated with argon at high temperatures. In this geometry, the initial layers of $^4\text{He}$ form solid phases adjacent to the walls, while the final layer forms a fluid core at the center of the pore as confirmed by numerical simulations \cite{kulchytskyy2013local, Nichols:2020of}.  We investigated this core liquid for pore fillings ranging from \SIrange{10.51}{15}{\milli\mol\per\gram} and our measurements reveal signatures of emergent one-dimensional behavior. Notably, we find that the excitations of the confined quantum liquid show no dependence on pore filling. We interpret this as indicating that the core liquid forms in clustered regions rather than a uniform dilute phase. Furthermore, temperature-dependent measurements of the TLL response yield an activation energy of \SI{19}{\kelvin}.



\section{Experimental Procedure}\label{sec3}


Our sample was obtained from Sigma-Aldrich \cite{mcm41SA:2008} and was characterized using X-ray powder diffraction and N$_2$ gas adsorption isotherms \cite{Nichols:2020of,Soni:2026ie}. The sample consisted of a single phase with pores arranged on a hexagonal lattice with a lattice constant of \SI{47}{\angstrom}. The MCM-41 had a surface area \cite{Brunauer:1938pz} of \SI{915}{\meter\squared\per\gram} and a pore diameter \cite{Jaroniec:1999mi} of \SI[uncertainty-mode=separate]{30.0(30)}{\angstrom}. Ar isotherms at \SI{90}{\kelvin} yielded a monolayer coverage of \SI{8.994}{\milli\mole\per\gram}. Once a monolayer of Ar was deposited on the surface, the sample was slowly cooled to \SI{4.2}{\kelvin}, yielding a confining medium with a pore diameter of $\sim$\SI{20}{\angstrom}.

We performed inelastic neutron scattering measurements using the Cold Neutron Chopper Spectrometer (CNCS) at the Spallation Neutron Source\cite{CNCS}.  CNCS is a direct geometry time-of-flight spectrometer which views a cold coupled moderator.  A high-speed Fermi chopper is used to select the energy of the incident neutron pulse. The speeds and slit widths of the choppers can be varied, allowing adjustments to the incident energy and  resolution. The final flight path consists of 400 $^3$He position sensitive detectors \SI{3.5}{\meter} from the sample position.  The detector covers scattering angles from \SIrange{-50}{140}{\degree} horizontally and $\pm\SI{16}{\degree}$ vertically.

Measurements were carried out with incident energies of \SIlist{6;12;28}{\milli\eV} which allowed maximum momentum transfers of \SIlist{3;4.5;6}{\angstrom^{-1}}, respectively. CNCS has three standard operating modes: High resolution (HR), Intermediate resolutions (AI) and High flux (HF). Only the HR and AI modes were used for these studies.  Measurements at \SI{6}{\milli\eV} using the HR provided a resolution of \SI{0.138}{\milli\eV} and the AI mode provided a resolution of \SI{0.190}{\milli\eV}.  Measurements at \SI{12}{\milli\eV} provided a resolution of 0.340 meV in the HR mode and 0.410 meV using the AI MODE.  Measurements at 28 meV utilized HR mode provided resolution of 0.88 meV and 1.30 meV with the AI mode Standard data reduction routines \cite{DAVE} are used to convert the observed scattering to the dynamic structure factor $S(Q,E)$ where $Q$ is the transferred momentum and $E$ the transferred energy.

The sample for these studies consisted of \SI{6.13}{\gram}.  The MCM-41 was in the form of cylindrical pellets \SI{1.25}{\centi\meter} in diameter and \SI{1}{\centi\meter} high with a mass of \SI{0.875}{\gram}.  The pellets were baked in vacuum at \SI{120}{\celsius} to remove adsorbed water vapor.  The sample was then transferred to an aluminum sample cell in a nitrogen glove box.  The cell was a cylindrical aluminum can of outer diameter \SI{1.5}{\centi\meter}, a height of \SI{6}{\centi\meter}, and a wall thickness of \SI{1}{\milli\meter}.  The pellets were separated by cadmium spacers to reduce multiple scattering. A top-loading liquid helium cryostat with aluminum tails, commonly referred to as an ``orange" cryostat, was used to obtain the low temperatures examined in this study. A silicon diode was used to monitor the sample temperature.


\section{Results}\label{sec3}


We carried out measurements with fillings of \SIlist{10.51;11.56;12.1;12.77;12.87;13.5;14.06;15.00}{\milli\mol\per\gram}.  No significant inelastic scattering, characteristic of the TLL, was observed for the lowest \SI{10.51}{\milli\mol\per\gram} filling.  This is consistent with the formation of strongly bound layers at the pore surface, as these layers would be strongly localized and exhibit little inelastic scattering.
This is consistent with our observation, described later, that the emergent quantum liquid does not appear until $\approx \SI{11}{\milli\mol\per\gram}$.

The results for \SI{14.06}{\milli\mol\per\gram} at \SI{1.5}{\kelvin} are shown in Fig.~{\ref{SQE}} for three different incident energies using the AI mode. The incident energy determines both the maximum momentum transfer ($Q$) that can be reached and the instrumental resolution. The intense scattering near $E=\SI{0}{\milli\eV}$ is due to scattering from the MCM-41 matrix, the argon monolayer, and the immobile helium layers. The width of this elastic scattering is a measure of the instrumental resolution.  As can be seen, the \SI{6}{\milli\eV} data has a resolution of $\simeq \SI{0.2}{\milli\eV}$ while the \SI{28}{\milli\eV} data has a resolution of $\simeq$ \SI{1.5}{\milli\eV}.

For comparison, the green curve shows the bulk $^4$He excitation dispersion with its characteristic phonon-maxon-roton structure \cite{henshaw1961modes,cowley1971inelastic,Glyde:2017es,Beauvois:2018pm}.  The excitation energy initially increases linearly with $Q$ until reaching a peak at the maxon.  The scattering then decreases to a minimum value at the roton at $Q_r \simeq \SI{1.9}{\angstrom^{-1}}$ with an energy gap of $\sim \SI{0.75}{\milli\eV}$.  Beyond the roton, the excitation energy increases until it plateaus due to two-roton scattering. The excitations in liquid helium are well defined so the scattering intensity is well localized and there is essentially no scattering intensity as one moves off the excitation energies.

The scattering shown in Fig.~{\ref{SQE}} makes it quite clear that we are not observing bulk liquid $^4$He.  In particular, at \SI{6}{\milli\eV} there is significant scattering intensity below the roton minimum.  Bulk $^4$He would have no scattering intensity below \SI{0.75}{\milli\eV}, whereas it is clear from the measurements that there is significant scattering intensity down to the resolution limit of $\simeq \SI{0.2}{\milli\eV}$. 

    Neutron scattering probes density fluctuations in the confined liquid.  For a single pore, let $Q$ denote the component of the transferred wavevector along the pore axis. At small enough $Q$, the linear TLL theory describes these fluctuations as sound waves with energy $E=\hbar v|Q|$. At larger wavevectors, the curvature of the excitation spectrum becomes important and the density response can extend over a continuum of energies with a momentum-dependent threshold $E_{\rm th}(Q)$ setting the smallest energy for excitations to occur.  Nonlinear TLL theory describes the response near its lower boundary in terms of a mobile hole coupled to the long-wavelength density waves of the liquid \cite{ Pustilnik:2006bs,Pereira:2008bw,Imambekov:2009ly,Imambekov:2009pr,Imambekov:2012ho}. Here, it is important to note that the hole is an effective excitation of the interacting helium, rather than an additional impurity atom.

At zero temperature, the response close to this boundary has the form (using the notation of Ref.~\cite{Motta:2016ge})
\begin{equation}
    S_{\rm 1D}(Q,E) = \Theta\!\left(E-E_{\rm th}(Q_n^\ast)\right)
    \left|E-E_{\rm th}(Q_n^\ast)\right|^{-\mu(Q)},
    \label{eq:1DSQE}
\end{equation}
assuming Galilean invariance where $Q_n^\ast = Q - 2n k_{\rm F}$.  The exponent $\mu(Q)$ describes how the scattering intensity approaches the threshold: positive values produce an enhancement, negative values give an intensity that vanishes at the boundary, and $\mu(Q)=0$ gives a step. The exponent is non-universal, depending on both wavevector and interactions in general, and can be determined exactly in terms of phase shifts for some exactly solvable models. 

A useful model to understand the effects of strong short-range repulsion (as is present in the hard-core repulsion between two $^4$He atoms) is a one-dimensional fluid of impenetrable hard rods \cite{Motta:2016ge,Bertaina:2016gu}. For a uniform fluid with linear number density $\rho_{\rm 1D}$, we define $k_{\rm F}=\pi\rho_{\rm 1D}$ and $E_{\rm F}=\hbar^2 k_{\rm F}^2/(2m)$, where $m$ is the mass of a $^4$He atom. These quantities provide convenient wavevector and energy scales and reflect the fact that impenetrable bosons and the corresponding spinless fermions have the same density correlations in 1D \cite{Giamarchi2004}.

Within the hard-rod approximation for $^4$He, the excitation threshold is given by:
\begin{equation}
    E_{\rm th}(Q_n^\ast) \simeq
    {4E_{\rm F}}{K_L}
    \left[
        \frac{Q_n^\ast}{2k_{\rm F}}
        -\left(\frac{Q_n^\ast}{2k_{\rm F}}\right)^2
    \right]\, ,
    \label{eq:Ethreshold}
\end{equation}
where $k_{\rm F}$ sets the density scale, and $K_L$ controls the energy scale at fixed density. 

The orange line in Fig.~{\ref{SQE}} shows a fit of the theoretical prediction for the threshold of a purely 1D quantum liquid of hard spheres with a Fermi wavevector of $k_{\rm F}$ $\approx$ 0.8 \AA$^{-1}$ and a Luttinger parameter of $K_L \approx 1$.  As can be seen, the hard sphere model qualitatively captures the behavior of the observed scattering.  

The Fermi momentum and Luttinger parameter were obtained by fitting the observed scattering to Eq.~\eqref{eq:1DSQE}. The width of the observed scattering is primarily determined by the instrumental resolution in both $E$ and $Q$.  Therefore, $\mu(Q)$ was simply set equal to zero and the instrumental resolution was represented by a Gaussian.

Typically, the excitation energy spectrum would be obtained by taking a cut of $S(Q,E)$ at constant $Q$.  The resulting scattering, $S(Q=\text{const.}, E)$, would then be fit to a Gaussian to obtain the excitation energy in energy. Repeating this for different values of $Q$ then allows $E_{\rm th}(Q)$ to be determined.

Unfortunately, the excitation energy is changing rapidly with $Q$,  thus, $S(Q=\text{const.}, E)$ is quite spread out, making extraction of accurate values for the maximum scattering difficult. To address this, we have taken a cut at a constant energy and fit $S(Q,E=\text{const.})$ to obtain the momentum associated with this energy.  The resultant $E_{\rm th}(Q)$ were fit to Fig.~{\ref{EvsQ} to obtain $K_L$ and $k_{\rm F}$.

    The results for fillings between $11.56$ and \SI{15.0}{\milli\mol\per\gram} are shown in Fig.~{\ref{Kl_fill}}. In Ref.~\cite{del2022experimental},  Del Maestro et. al. reported values at \SI{12.1}{\milli\mol\per\gram} of $K_L=1.18^{+0.38}_{-0.2}$ and $k_{\rm F}=0.80^{+0.02}_{-0.02}$, which are consistent with these results.  Surprisingly, the extracted values for both the Luttinger parameter and the Fermi momentum do not exhibit any systematic variation with filling.  Under the assumption that these values are density independant we find that $K_L=1.056^{+0.147}_{-0.079}$ and $k_{\rm F}=0.82^{+0.01}_{-0.01}$.

    The relative intensity of the scattering at different fillings was also analyzed. The relative intensity was determined by calculating the total integrated scattering for $\SI{0.5}{\milli\eV} <E< \SI{1.0}{\milli\eV}$ and $\SI{1.5}{\angstrom^{-1}} <Q< \SI{3}{\angstrom^{-1}}$.  While the Luttinger parameter and the Fermi momentum showed no systematic variation with filling, the relative intensity, shown in Fig.~{\ref{I_fill}} varied systematically with filling.  Assuming that all helium added above a threshold filling of $\approx \SI{11}{\milli\mol\per\gram}$ goes into the core liquid,  then the scattering intensity should be proportional to the filling. The results are well described by a simple linear relationship with a slope of $0.38 \pm 0.048$ and intercept $-4.29\pm0.62$. The threshold for the appearance of the 1D liquid is then predicted to be \SI{11.29(0.24)}{\milli\gram\per\mol}.

We also studied the temperature dependence of the scattering at a filling of \SI{13.51}{\milli\mol\per\gram}. As the temperature was increased there was no statistically significant change in either $K_L$ or $k_{\rm F}$.  However, the overall intensity of the scattering decreased with increasing temperature as shown in Fig.~{\ref{T_Dep}}.  The observed temperature dependence could be fit to a thermally activated behavior
\begin{equation}
    I(T)=I(0)-f\exp(-\Delta/T)
    \label{eq:Activation}
\end{equation}
where $I(0)$ is the scattering intensity at $T=0$ and $\Delta$ is the activation energy. Fitting this expression to the observed temperature dependence yields $I(0)=0.989\pm0.015$, $f=-5.91\pm5.1$, and $\Delta=\SI{19.1(5.0)}{\kelvin}$. This would indicate that the TLL excitations would disappear at $\approx 10 K$.


\section{Conclusion}


The initial goal of this study was to probe whether Ar preplated MCM-41 was a viable confinement platform where one-dimensional $^4$He could be studied as a function of density from the weakly interacting to strongly interacting limit. Theoretical studies had indicated that as helium was added to the system, three highly localized layers would develop and then additional helium would be formed as a 1D liquid at the center of the pore.  For a limited range of external pressures, the density of the core liquid would evolve as more $^4$He was added \cite{Nichols:2020of}.  In this view, the helium in the core liquid would be uniformly spaced as illustrated in Fig.~{\ref{linear}}.  In this model both $K_L$ and $k_{\rm F}$ would be predicted to evolve smoothly with filling. 


The filling-independent excitation parameters and increasing spectral weight are consistent with an increasing fraction of the pore cores occupied by liquid at approximately fixed local density. One possible realization is the growth of separated liquid segments, although the present measurements do not distinguish segment growth from the formation of additional segments or the filling of additional pores. 

The temperature dependence of the scattering also lends support to the development of a TLL phase. As the temperature is increased, the scattering exhibits TLL behavior with no statistically significant change in $K_L$ or $k_{\rm F}$.  The only change is in the intensity of the scattering, which decreases by $\approx$ 30\% in going from \SIrange{1.5}{6}{\kelvin}. It is important to note that TLL behavior is observed at the highest temperature studied, which is approximately three times the transition temperature for superfluid $^4$He supporting the identification of this phase as a TLL.  The temperature dependence of the scattering was fit to a thermally activated behavior with a characteristic temperature of \SI{19}{\kelvin}.  The characteristic temperature where we expect TLL behavior to be observed is $T \ll \hbar v Q/k_{\rm B} \simeq 4E_{\rm F}/(K k_{\rm B}) \simeq 13{K}$ for $Q \simeq 2k_{\rm F}$ where $k_{\rm B}$ is Boltzmann's constant.\\


\bmhead{Acknowledgements}

This research was supported by the Department of Energy (DOE) under BES Award WS01054108.


\begin{figure}[h]
\centering
\includegraphics[width=0.8\textwidth]{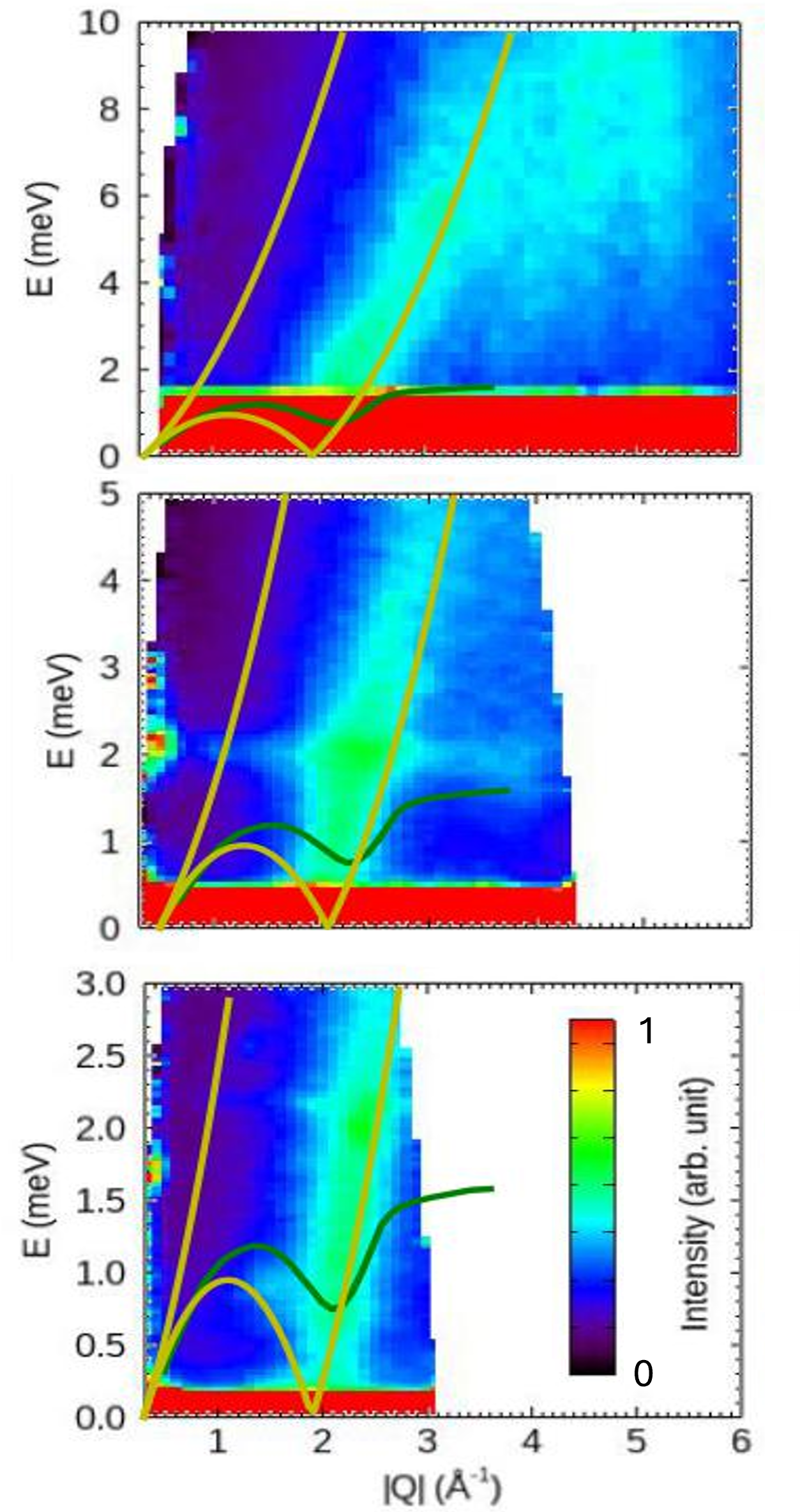}
\caption{The dynamic structure factor, $S(Q,E)$, from helium confined in the Ar-preplated pores of MCM-41.  The scattering was measured at a filling of \SI{13.5}{\milli\mol\per\gram}. Top: 28 meV incident energy. Middle: 12 meV Incident energy. Bottom: 6 meV incident energy}
\label{SQE}
\end{figure}

\begin{figure}[h]
\centering
\includegraphics[width=0.8\textwidth]{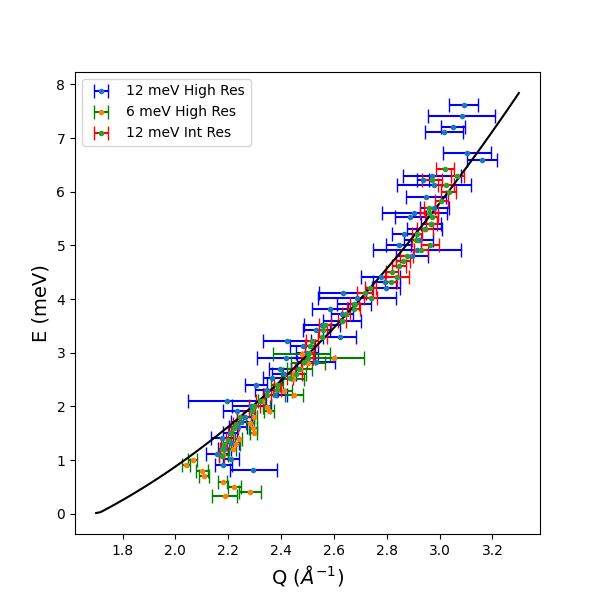}
\caption{The peak excitation energies as a function $Q$ at a filling of \SI{11.56}{\milli\mol\per\gram} for data at 6, 12, and 28 meV in the HR mode.  Solid curve shows the fit of equation~\eqref{eq:Ethreshold} to the data with .}
\label{EvsQ}
\end{figure}

\begin{figure}[h]
\centering
\includegraphics[width=0.5\textwidth]{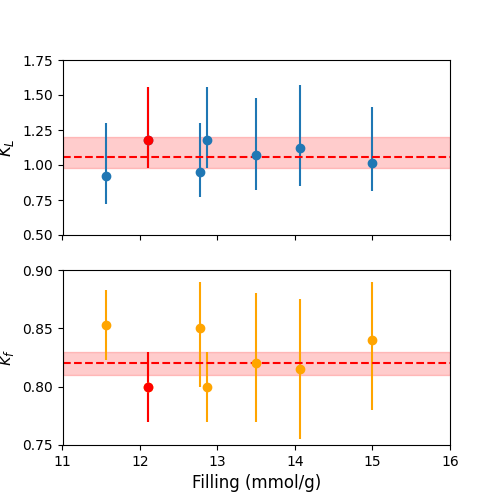}
\caption{The Luttinger liquid parameter (top) and Fermi momentum (bottom) versus filling. The red dashed lines show the average values of $K_L$ and $k_F$, while the pink shaded region shows the error band. }
\label{Kl_fill}
\end{figure}

\begin{figure}[h]
\centering
\includegraphics[width=0.5\textwidth]{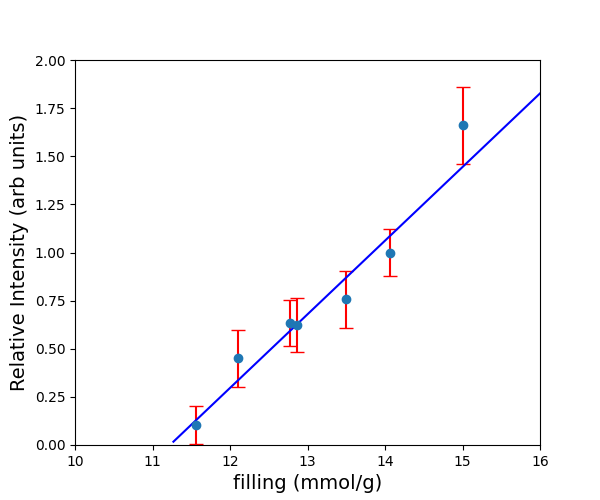}
\caption{Inelastic scattering intensity, I, as a function of filling, f. The straight line is a fit of the data to I=(\SI{0.38}{\gram\per\milli\mol})f-4.29}\label{I_fill}
\end{figure}

\begin{figure}[h]
\centering
\includegraphics[width=0.5\textwidth]{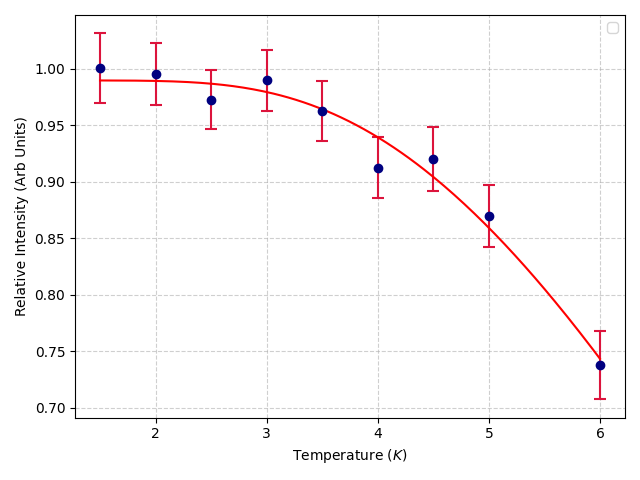}
\caption{The symbols are the measured temperature dependence of the excitations at a filling of \SI{13.51}{\milli\mol\per\gram}.  The curve is a fit to a thermally activated temperature dependence.}
\label{T_Dep}
\end{figure}

\begin{figure}[h]
\centering
\includegraphics[width=0.5\textwidth]{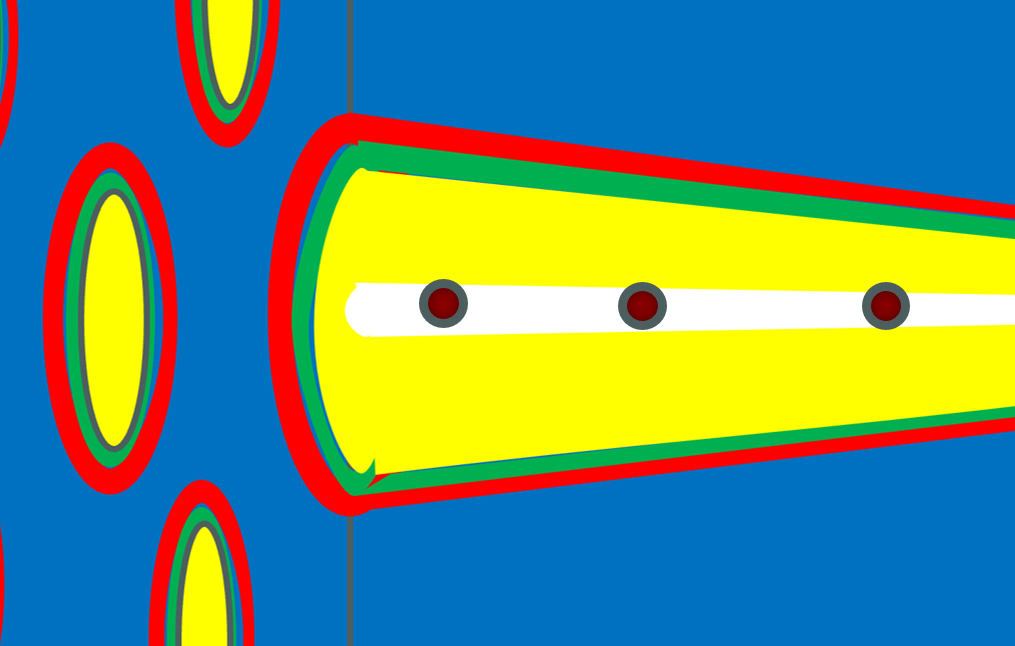}
\caption{Illustration of 1D helium uniformly distributed.  The blue represents the MCM-41 matrix with it array of hexagonally spaced 1D pores.  The green represents the monolayer of argon on the pore surface.  The yellow represents the solid layers of the helium.  The dots represent the $^4$He atoms at the pore center.}
\label{linear}
\end{figure}

\begin{figure}[h]
\centering
\includegraphics[width=0.5\textwidth]{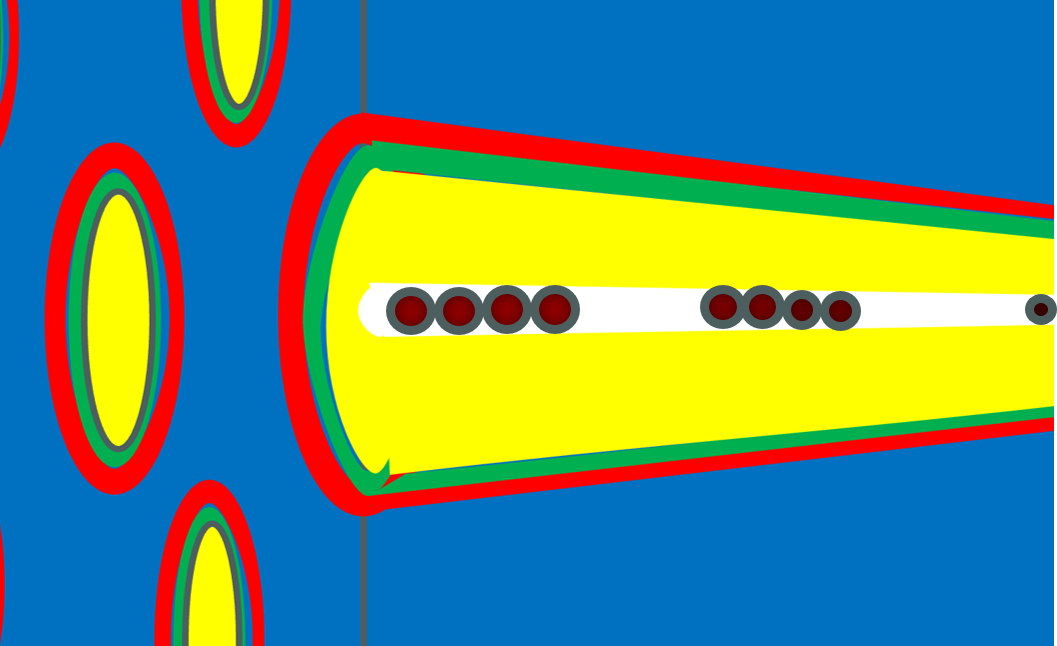}
\caption{Illustration of 1D helium forming dense islands at the pore center. The color coding is the same as Fig.~{\ref{linear}}}.
\label{cluster}
\end{figure}


\bibliography{sn-bibliography}

\end{document}